\documentclass{llncs}
    \usepackage{lineno}

\DeclareMathAlphabet{\mathsl}{OT1}{cmr}{m}{sl}

\newcommand\ie{\hbox{\textit{i.e.}}}

\newcommand{\false}{\texttt{f}}

\newcommand{\true}{\texttt{t}}

\newcommand{\defsymboldelta}{\stackrel{\Delta}  {=}}

\newcommand\Fifex[6]{#1;#2\mathrel{\buildrel #3\over{\hbox to #6pt{\rightarrowfill}}}#4;#5}

\newcommand\ra\rightarrow

\newcommand\tensor\otimes

\newcommand{\gdash}{\vdash\kern -4pt\vdash}

\newcount\PLv\newcount\PLw\newcount\PLx\newcount\PLy\newdimen\PLyy\newdimen\PLX
\newbox\PLdot \setbox\PLdot\hbox{\tiny.} \def\scl{.08} 
\def\PLot#1{\PLx`#1\advance\PLx-42\PLy\PLx\PLv\PLx\divide\PLy9\PLw\PLy\multiply
\PLw9\advance\PLx-\PLw\advance\PLx-4\PLy-\PLy\advance\PLy4\PLX=\the\PLx pt
\advance\PLyy\the\PLy pt\wd\PLdot=\scl\PLX\raise\scl\PLyy\copy\PLdot}
\def\draw#1{\ifx#1\end\let\next=\relax\else\PLot#1\let\next=\draw\fi\next}

\def\invamp{\hbox{\PLyy=70pt\draw :::;DMV_gqppyyyyyooooxxxnnwvlutkjaWNE=5-./9
9:::CCCC:::99/..--544=EENWWaajjjkktttttttNNNVVVVVVVV\end \hskip4pt}}
\newbox\iabox\setbox\iabox\invamp

\long\def\hide#1\endhide{}

\DeclareMathAlphabet{\mathsl}{OT1}{cmr}{m}{sl}

\newcommand{\rTell}{\rm R_{TELL}}

\newcommand{\rSum}{\rm R_{SUM}}
\newcommand{\rSumCS}{\rm R'_{SUM}}

\newcommand{\rLocal}{\rm R_{LOC}}

\newcommand{\rCall}{\rm R_{CALL}}

\newcommand{\rEquiv}{\rm R_{EQUIV}}

\newcommand{\vx}{\overline{x}}

\newcommand{\vy}{\overline{y}}
\newcommand{\vt}{\overline{t}}

\newcommand{\Barb}[2]{#1\Downarrow_{#2}}
\newcommand{\BarbCLP}[2]{#1\Downarrow_{#2}^{\cH}}

\def\cC{\mathcal{C}}
\def\cD{\mathcal{D}}

\def\cH{\mathcal{H}}

\newcommand{\tellp}[1]{\mathbf{tell}(#1)}

\newcommand{\whenp}[2]{\mathbf{ask} \  (#1) \  \mathbf{then} \ #2}

\newcommand{\nextp}[1]{ \mathbf{next}\ #1}

\newcommand{\localp}[2]{(\mathbf{local} \, #1)  \, #2}

\newcommand{\skipp}{\mathbf{skip}}

\newcommand{\defsymbol}{\stackrel{\Delta}  {=}}

\newcommand{\ccp}{CCP}

\newcommand{\os}{[\![}
\newcommand{\cs}{]\!]}

\newcommand{\equivP}{\cong}

\newcommand{\entails}{\models}

\long\def\comment#1{}

\newcommand{\redi}{\longrightarrow}
\newcommand{\rediCLPP}[1]{\longrightarrow_{CLP(#1)}}
\newcommand{\rediCLP}{\rediCLPP{\cH}}
\newcommand{\redirex}{\longrightarrow^{*}}

\newcommand{\rediIdxJ}[2]{\xrightarrow{[#1]_{#2}}}

\newcommand{\sliced}{\bullet}

\newcommand{\stopc}{\texttt{stop}}
\newcommand{\beatc}{\texttt{beat}}
\newcommand{\Mdmtwo}{\texttt{Mdm2}}
\newcommand{\MdmtwoA}{\texttt{Mdm2A}}

\newcommand{\past}[1]{\circleddash#1}

\newcommand{\idxP}[1]{\!:\!#1}

\newcommand{\lr}[1]{\langle #1 \rangle}
\newcommand{\predAssertionA}[2]{#1[#2]}
\newcommand{\predAssertionE}[2]{#1\langle#2\rangle}
\newcommand{\fProc}[1]{procs(#1)}
\newcommand{\fStore}[1]{store(#1)}
\newcommand{\fSymp}[1]{symp(#1)}
\usepackage{subfig}
\usepackage{fancyvrb}
\usepackage{graphicx}
\usepackage{wrapfig}
\usepackage{todonotes}
\usepackage{times}

\usepackage{color}
\usepackage{latexsym}
\usepackage{amssymb}
\usepackage{amsmath}
\usepackage{proof} 
\usepackage{listings}
\usepackage{lineno}

\usepackage{url}
\usepackage{verbatim}
\usepackage[linesnumbered]{algorithm2e}

\newcommand{\posC}[1]{\texttt{pos}(#1)}
\newcommand{\negC}[1]{\texttt{neg}(#1)}
\newcommand{\consC}[1]{\texttt{cons}(#1)}
\newcommand{\iconsC}[1]{\texttt{icons}(#1)}

\newcommand{\entailsF}{\entails_{\mathcal{F}}}
\newcommand{\postF}[1]{\texttt{post}(#1)}
\newcommand{\invF}[1]{\texttt{inv}(#1)}

\begin{document}
\long\def\comment#1{}

\title{An assertion language for slicing\\ constraint logic languages}

\author{ Moreno   Falaschi\inst{1} \and Carlos Olarte\inst{2}}
\institute{Dept. Information Engineering and Mathematics, 
              Universit\`a di Siena, Italy.\\
		      \email{moreno.falaschi@unisi.it.}
	 \and
	 ECT, Universidade Federal do Rio Grande do Norte, Brazil\\
	 \email{carlos.olarte@gmail.com.}
	   }

	   \date{}
\maketitle

\begin{abstract}
Constraint Logic Programming (CLP) is a language scheme for combining
two declarative paradigms: constraint solving and logic programming.
Concurrent Constraint Programming (\ccp) is a declarative model for
concurrency where agents interact by telling and asking constraints
 in a shared store.
In a previous paper, we  developed a framework for dynamic slicing 
of \ccp\ where the user first  identifies that a (partial) computation is wrong. 
Then, she marks (selects) some parts of the final state
corresponding 
to the data (constraints) and processes  that she wants to study
more deeply. An automatic process of slicing begins,
and the partial computation is ``depurated''  by removing irrelevant 
information. 
In this paper we give two major contributions. First,
we extend the framework to CLP, thus
generalizing the previous work. 
Second, we provide an assertion
language suitable for both, CCP and CLP, 
which allows the user to specify some properties
of the computations in her program. If a state in a 
computation does not satisfy
an assertion then some ``wrong'' information is identified
and an automatic slicing process can start. 
We thus make one step further towards
automatizing the slicing process. We show that our
framework can be integrated with the previous semi-automatic one,
giving the user more choices and flexibility. We show by means
of examples and
experiments the usefulness of our approach.
\end{abstract}
 

\section{Introduction}\label{sectionintroduction}

Constraint Logic Programming (CLP) is a language scheme 
\cite{DBLP:journals/jlp/JaffarMMS98} 
for combining two declarative paradigms: constraint solving 
and logic programming (see an overview in \cite{JM94}).
Concurrent Constraint Programming (CCP)  \cite{saraswat91popl}  
(see a survey in \cite{DBLP:journals/constraints/OlarteRV13}) 
combines   concurrency primitives with the ability 
to deal with constraints, and hence, with partial information. 
The notion of concurrency 
is based upon the shared-variables communication model.
CCP is intended for reasoning, modeling and 
programming 
concurrent  agents (or processes) that interact with each other and
their environment by posting and asking information in a medium, 
a so-called \emph{store}.
CCP is a very flexible model and 
has been applied to an increasing number of different 
fields such as probabilistic and stochastic, timed  and mobile  
systems~\cite{Olarte:08:SAC,Br11},
and more recently to social networks with spatial and epistemic 
behaviors \cite{DBLP:journals/constraints/OlarteRV13}, as well
as modeling of biological systems~\cite{CFOP10,CFHOT15,OCHF16,BBDFH18}. 

One crucial problem with constraint logic languages is to define
appropriate debugging tools.
Various techniques and several frameworks have been proposed for debugging 
these languages.
Abstract interpretation techniques have been 
considered (e.g. in \cite{CFM94,CominiTV11absdiag,FOP09,DBLP:journals/tplp/FalaschiOP15}) 
as well as (abstract) declarative debuggers following the seminal work of 
Shapiro \cite{Shapiro83}.
However, these techniques 
are approximated (case of abstract interpretation) or it can be difficult 
to apply them when dealing with complex programs (case of 
declarative debugging) as the user should answer to too many
questions. 

In this paper we follow a technique inspired by slicing.
Slicing was introduced in some pioneer works by Mark Weiser \cite{MW84}. 
It was originally defined as a static technique, independent of  any particular 
input of the program. 
Then, the technique was extended by introducing the so called 
dynamic program slicing \cite{KL88}. This technique
is useful for simplifying the debugging process, by selecting a portion of
the program containing 
the faulty code.
Dynamic program slicing has been applied
to several programming paradigms (see \cite{Silva2012} for a  survey). 
In the context of constraint logic languages, we defined a 
tool \cite{FGOP2016} able to interact with 
the user and filter, in a given computation,  the information which is 
relevant to a particular observation or result.
In other words, the programmer could mark (select) the information 
(constraints, agents or atoms) that she is 
interested to check in a particular  computation that she 
suspects to be wrong. Then, a corresponding 
depurated partial computation is obtained automatically, 
where only the information relevant to
the marked parts is present.

In a previous paper \cite{FGOP2016} we 
presented the first formal framework for debugging CCP 
via dynamic slicing. 
In this paper we give two major contributions.
First, we extend our framework
to  CLP. Second, we introduce an assertion language
which is integrated within the slicing
process for automatizing it further.
The extension to CLP is not immediate,
as while for CCP programs non-deterministic choices give  rise to one single
computation, in CLP all computations corresponding to different non-deterministic choices can be followed and can lead to different
solutions. Hence, some rethinking of the the framework is necessary.
We show that it is possible to define a transformation
from CLP programs to CCP programs, which allows us to show that
the set of observables of a CLP program and of its
translation to a CCP program correspond.
This result also shows that the 
computations in the two languages are pretty similar and 
the
framework for CCP can be extended to deal with CLP programs.

Our framework \cite{FGOP2016} consists of three main steps.
First the standard operational semantics of the sliced language is 
extended to an  enriched semantics 
that adds to the standard semantics the needed meta-information 
for  the slicer. 
Second, we consider several analyses of the faulty 
situation based on the program wrong behavior,
including causality, variable dependencies, unexpected behaviors and 
store inconsistencies. 
This second step was left to the user's responsibility: the user
had to examine the final state of the faulty computation and 
  manually  mark/select  
  a subset of  constraints that she wants to study further.
The third step  
is an automatic marking algorithm
that removes the information not relevant to derive the 
constraints selected in the second step. 
This algorithm is flexible and applicable to timed extensions of 
\ccp\ \cite{DBLP:journals/jsc/SaraswatJG96}. Here, for CLP programs we introduce also the possibility to mark 
atoms, besides constraints.

We believe that the second step above, namely 
identifying the right state and the relevant information 
to be marked, can be difficult for the user and we believe that
it is possible to improve automatization of this step.
Hence, one
major contribution of this paper is to introduce a 
specialized assertion language
which allows the user to state properties of the computations
in her program. If a state in a 
computation does not satisfy
an assertion then some ``wrong'' information is identified
and an automatic slicing process can start.
We show that assertions can be integrated in our previous semi-automatic 
framework \cite{FGOP2016},
giving the user more choices and flexibility. 
The assertion language is a good companion to the already implemented 
tool for the slicing of \ccp\ programs to automatically detect 
(possibly) wrong behaviors  
and stop the computation when needed. The framework 
can also be applied to timed variants of CCP. \\

\paragraph{Organization and Contributions } Section \ref{sectionccp} 
describes CCP and CLP and their  operational semantics.
We   introduce a translation
from CLP to CCP programs and prove a correspondence theorem
between successful computations.
In Section \ref{sectionslicing} we recall the slicing 
technique for CCP  \cite{FGOP2016} and extend it to CLP. 
The extension of our framework to CLP is our first 
contribution.
As a second major contribution, in Section \ref{sectionassertions} we present our 
specialized assertion language and describe its main operators and
functionalities.
In Section   \ref{sec:ex} we show some examples to 
illustrate the expressiveness of our extension, and 
the integration into the former tool.
Within our examples we show how to automatically debug 
a biochemical system specified in timed CCP and one classical
search problem in CLP. 
Finally, Section \ref{sectionconclusions} discusses some related work
and concludes.

\section{Constraint Logic Languages}\label{sectionccp}

In this section we  define an operational semantics 
suitable for both,  CLP 
~\cite{JM94} 
and \ccp\ programs  \cite{saraswat91popl}.
We start by defining \ccp\ programs and then we  obtain CLP 
 by restricting the set of  \ccp\ operators. 

Processes in \ccp\ \emph{interact} with each other by \emph{telling} and 
\emph{asking} 
constraints (pieces of information) in a common 
store of partial information. The type of constraints
 is not fixed but parametric in a constraint system (CS), a central notion for both \ccp\ and CLP. 
 Intuitively, a
CS  provides a signature from which  constraints
can be built from basic tokens (e.g., predicate symbols), and two basic 
operations: conjunction $\sqcup$ (e.g., $x\neq y \sqcup x > 5$) 
and variable hiding $\exists$ (e.g., $\exists x . y = f(x)$). 
As usual, $\exists x. c$ binds $x$ in $c$. 
The CS defines also an
\emph{entailment} relation ($\entails$) specifying inter-dependencies
between constraints:  $c\entails d$  means that the
information $d$ can be deduced from the information 
$c$ (e.g., $x>42 \entails x>37$). 
We shall use $\cC$ to denote the set of constraints with typical elements 
$c,c',d,d'...$. We assume that there exist $\true, \false \in \cC$, such that
for any $c \in \cC$, $c \entails \true$ and $\false \entails \cC$.
The reader 
may refer to
 \cite{DBLP:journals/constraints/OlarteRV13}
  for different formalizations and examples of constraint systems.

\paragraph{\bf The language of \ccp\ processes.}
In process calculi, the language of processes in  
\ccp\  is given by a small number of 
primitive operators or combinators. 
 Processes are built from constraints in the underlying constraint 
 system and the following syntax:

$
P,Q ::= \skipp\mid \tellp{c} \mid   
\sum\limits_{i\in I}\whenp{c_i}{P_i} \mid   
P \parallel Q  \mid 
\localp{x}{P}   \mid  
p(\overline{x})
$

The process $\skipp$  represents inaction. The process 
 $\tellp{c}$ adds $c$ to the current store $d$ producing the new store $c\sqcup d$.
 Given a non-empty finite set of indexes $I$, the process $\sum\limits_{i\in I}\whenp{c_i}{P_i}$ non-deterministically chooses  $P_k$ for execution if the   store entails $c_k$. The chosen alternative, if any, precludes the others.
This provides a powerful synchronization mechanism based on constraint entailment.   
 When $I$ is a singleton, we shall omit the ``$\sum$'' and we simply write $\whenp{c}{P}$. 
 
The process $P\parallel Q$ represents the parallel 
(interleaved) execution of $P$ and $Q$. 
The process 
$\localp {x}{P}$ behaves as $P$ and binds the variable 
$x$ to be local to it.

Given a process definition  $p(\overline{y}) \defsymboldelta P$,  
where all free variables of $P$ are in the set of pairwise distinct
variables $\overline{y}$, the process $p(\overline{x})$  evolves into 
$P[\overline{x}/\overline{y}]$.  A \ccp\ program takes the form 
$\cD.P$ where $\cD$ is  a set of  process definitions and $P$ is a process.

The Structural Operational Semantics (SOS)   of \ccp\ 
is given by the transition relation 
$ \gamma \redi \gamma'$  
satisfying the rules in Figure \ref{fig:sos}.
Here we follow the formulation 
 in \cite{fages01ic} where the local variables created by the program appear explicitly in 
 the transition system and parallel composition of agents is 
 identified by a multiset of agents. 
More precisely, a \emph{configuration} $\gamma$ is a triple of the  form  
$(X;  \Gamma ;  c)$, where $c$ is a constraint representing the  store,  $\Gamma$ is a multiset of 
processes,
and $X$ is a set of hidden 
(local) variables of $c$ and $\Gamma$. 
The multiset $\Gamma=P_1,P_2,\ldots,P_n$  
represents the process  $P_1 \parallel P_2 \parallel \cdots \parallel P_n$. 
We shall indistinguishably
use both notations to denote parallel composition. Moreover, processes  
are quotiented by a structural 
congruence relation $\equivP$  satisfying: 
 (STR1) $P \equivP Q$ if $P$ and $Q$ differ only by a renaming of 
 bound variables (alpha conversion);
 (STR2) $P\parallel Q \equivP Q \parallel P$;
 (STR3) $P \parallel (Q \parallel R) \equivP (P \parallel Q)  \parallel R$; 
 (STR4) 
 $P \parallel \skipp \equivP P$.
  We denote by $\redirex$ the  reflexive and transitive closure of 
  a binary relation $\redi$.

\begin{definition}[Observables and traces]\label{def:obs}
A trace $\gamma_1\gamma_2\gamma_3\cdots $ is a sequence of configurations  
s.t. 
$\gamma_1 \redi \gamma_2 \redi \gamma_3 \cdots$. We shall use $\pi, \pi'$ 
to denote traces and $\pi(i)$ to denote
the i-th element in $\pi$. If 
$(X;\Gamma; d) \redirex(X';\Gamma';d')$ and 
$\exists X'. d' \entails c$ we write 
$\Barb{(X;\Gamma;d)}{c}$.
If $X=\emptyset$ and $d=\true$   we simply write   $\Barb{\Gamma}{c}$.
\end{definition}

Intuitively, if $P$ is a process then 
$\Barb{P}{c}$ says that $P$ can reach a store $d$ strong enough to entail $c$, \ie, $c$ is an output of $P$. 
Note that the variables in $X'$ above are hidden from $d'$  
since the information about  them is not observable.

\begin{figure}[t]
\resizebox{.95\textwidth}{!}
{
$
\begin{array}{ccc}
\infer[\rTell]
{(X; \tellp{c},\Gamma;d) \redi (X;\skipp, \Gamma;c\sqcup d)}
{}
\qquad
\infer[\rSum]{
  (X;\sum\limits_{i\in I}\whenp{c_i}{P_i},\Gamma;d) \redi (X;P_k,\Gamma;d)
  }
  {
  d \entails c_k \quad k\in I
  }
 \\\\
\infer[\rLocal]
{(X;\localp{x}{P},\Gamma;d) 
\redi (X\cup\{x\};P,\Gamma;d)
}
{x \notin X \cup fv(d) \cup fv(\Gamma)}
\qquad
\infer[\rCall]
{(X;p(\vx),\Gamma;d) \redi  (X;P[\vx/\vy],\Gamma;d) }
{p(\vy) \defsymboldelta P  \in \cD } \\\\
\infer[\rEquiv]
{(X;\Gamma;c) \redi (Y;\Delta;d)}
{(X;\Gamma;c) \equivP (X';\Gamma';c')  \redi (Y';\Delta';d') \equivP (Y;\Delta;d)}
\end{array}
$
}
\caption{Operational semantics for \ccp\ calculi\label{fig:sos}}
\vspace{-4mm}
\end{figure}

\subsection{The language of CLP}\label{sec:clp}

A CLP program \cite{DBLP:journals/jlp/JaffarMMS98} is a finite set of rules 
of the form 
\[
p(\vx) \leftarrow A_{1},\dots, A_{n}
\]
where  $A_{1},\dots A_{n}$, with $n\geq0$, are literals, i.e. 
either atoms or constraints in the underlying constraint 
system $\cC$,
and $p(\vx)$ is an atom. 
An atom has the form $p(t_{1},\ldots,t_{m})$, where $p$ is a user 
defined predicate symbol and the $t_{i}$ are 
terms from the constraint domain.

 The top-down operational semantics is given in terms of  derivations from goals 
 \cite{DBLP:journals/jlp/JaffarMMS98}. 
A configuration takes  the form $(\Gamma ; c)$ where $\Gamma$ (a 
goal) is a multiset of literals and $c$ is a constraint (the current store). 
The reduction relation is defined as follows. 

\begin{definition}[Semantics of CLP \cite{DBLP:journals/jlp/JaffarMMS98}]\label{def:sem-clp}
Let $\cH$ be a CLP program. 
A configuration $\gamma = (L_1,...,L_i,... L_n ; c)$ reduces to $\psi$, 
notation $\gamma \rediCLP \psi$, by selecting and removing a  literal $L_i$  
and then: 
\begin{enumerate}
 \item If $L_i$ is a constraint $d$ and $d \sqcup c \neq \false$, then 
 $\gamma \rediCLP (L_1,...,L_n ; c \sqcup d)$. 
 \item If $L_i$ is a constraint $d$ and $d \sqcup c = \false$ 
 (i.e., the conjunction of $c$ and $d$ is inconsistent), then 
 $\gamma \rediCLP (\Box ; \false)$ where $\Box$ represents the empty multiset of literals. 
 \item If $L_i$ is an atom  $p(t_1,...,t_k)$, then 
 $\gamma \rediCLP (L_1,...,L_{i-1},\Delta, L_{i+1}... , L_n ; c)$
 where one of the definitions for $p$, 
$p(s_1,...,s_k)\leftarrow A_{1},\dots, A_{n}
 $, 
 is selected and $\Delta = A_{1},\dots, A_{n}, s_1 = t_1 , ..., s_k = 
 t_k$. 
\end{enumerate}
A computation from a goal $G$ is a (possibly infinite) sequence 
$\gamma_1=(G ; \true)\rediCLP \gamma_2 \rediCLP \cdots$. 
We say that a computation finishes if the last configuration $\gamma_n$ cannot be reduced, 
i.e.,  $\gamma_n = (\Box ; c)$. In this case, if $c=\false$ then the derivation fails 
otherwise  we say that it succeeds. 
\end{definition}

Given a goal with free variables $\overline{x}=var(G)$,  
we shall also use the notation $\BarbCLP{G}{c}$ to denote that 
there is a successful computation $(G ;\true) \rediCLP^* (\Box;d)$ s.t. $\exists \vx. d \entails c$. 
We note that the  free variables of a goal  are progressively ``instantiated'' during computations by 
adding new constraints. 
Finally, the answers of a goal $G$,
notation $\BarbCLP{G}{}$ is the set
$\{\exists_{var(c)\backslash var(G)}(c) \mid (G ; \true) \rediCLP^* (\Box ; c) , c\neq \false \}$
where ``$\setminus$'' denotes set difference.

\paragraph{\bf From CLP to CCP. }
CCP  is a very general paradigm that  extends 
both
Concurrent  Logic Programming and Constraint Logic Programming \cite{NPV02}. 
However, in CLP,   we have to consider non-determinism of the type ``don't know''
\cite{Shapiro:1989}, which means that each predicate 
call can be reduced by using each rule which defines such a predicate.
This is different from the kind of non-determinism in \ccp, where the 
choice operator selects randomly one of the choices whose ask guard
is entailed by the constraints in the current store 
(see $\rSum$ in Figure \ref{fig:sos}). 

It turns out that by
restricting the syntax of CCP and giving an alternative interpretation to 
non-deterministic choices, we can have an encoding of CLP programs as CCP agents. 
More precisely, we shall remove the synchronization operator and we shall consider only 
blind choices of the form $Q = \sum\limits_{i\in I}\whenp{\true}{P_i}$. 
Note that $c\entails \true$ for any $c$ and then, the choices in the process $Q$ are 
not guarded/constrained. Hence, any of the $P_i$ can be executed regardless 
of the current store. This mimics the behavior of CLP predicates (see 
(3) in Definition \ref{def:sem-clp}), but with a different kind of 
non-determinism. The next definition formalizes this idea.

\begin{definition}[Translation]\label{def-trans}
Let $\mathcal{C}$ be a constraint system, $\mathcal{H}$ be a CLP program 
and $G$ be a goal. 
We define the set of \ccp\ process definitions $\os \mathcal{H} \cs = \cD$ as follows. For each 
user defined predicate symbol $p$ of arity $j$ and $1..m$ defined 
rules of the form
$
p(t^i_1,...,t^i_{j})\leftarrow  A^i_{1},\dots, A^i_{n_i}
$, 
we add to $\cD$ the following process definition 

\resizebox{.95\textwidth}{!}{
$
\begin{array}{lll}
p(x_1,...,x_j) &\defsymboldelta& \whenp{\true}{(\localp{~\overline{z_1}} {~\prod D_1 ~~ \parallel \os   A^1_{1}\cs ~\parallel  
\dots \parallel \os A^1_{n_1}~\cs}})  + ... + \\
& & \whenp{\true}{(\localp{\overline{z_m}}  {~\prod D_m\parallel  
 \os A^m_{1} \cs \parallel  \dots \parallel \os A^m_{n_m}\cs }})
\end{array}
$
}

where 
$\overline{z_i}= var(t^i_1,...,t^i_j) \cup var(A^i_1,...,A^i_{n_i})$, 
$D_i$ is the set of constraints $\{x_1 = t^i_1,...,x_j = t^i_j \}$,
$\prod D_i$
means $\tellp{x_1 = t^i_1} \parallel \cdots \parallel \tellp{x_j = t^i_j} $ and   literals are translated  as $\os A(\vt) \cs = A(\vt)$ (case of atoms) and $\os c \cs = \tellp{c}$ (case of constraints). 
Moreover, we translate the goal $\os A_1,...,A_n \cs $ as the process $ \os A_1 \cs \parallel \cdots \parallel 
\os A_n\cs$.

\end{definition}

We note that the head $p(\vx)$ of a process definition
$p(\vx) \defsymbol P$ in  \ccp\ 
can only have variables while a head  of a CLP rule  $p(\vt) \leftarrow B$
may have 
arbitrary terms with (free) variables. 
Moreover, in CLP, 
 each call to a predicate   returns a variant  with distinct 
 new variables (renaming the parameters of the predicate) \cite{DBLP:journals/jlp/JaffarMMS98}. 
 These two features of CLP can be encoded in CCP by first
 introducing local variables ($\localp{\vec{z_i}}{}$ in the above definition)
 and then, using constraints ($D_i$) to establish the connection between 
 the formal  and the actual parameters of the process definition. 
 
Consider for instance this simple CLP program dealing with lists: 
\begin{Verbatim}[fontsize=\scriptsize]
p([] , []) .
p([H1 | L1] , [H2 | L2]) :- c(H1,H2), p(L1,L2) .
\end{Verbatim}

and its translation 
\[
\begin{array}{lll}
p(x, y) &\defsymbol& \whenp{\true}{(\tellp{x=[]} \parallel \tellp{y=[]})} + \\
& & \whenp{\true}{\localp{X}{(\prod D \parallel c(H1,H2) \parallel p(L1, L2))}}
\end{array}
\]
where $D = \{x= [H1 | L1], y = [H2 | L2]\}$ and $X=\{H1,H2,L1,L2\}$.  
Note that   the CCP process  $p(l_a,l_b)$ can lead to 2 possible outcomes:
\begin{itemize}
 \item Using the first branch, 
 the  store becomes  $l_a=[] \sqcup l_b = []$.
 \item In the second branch, due to rule $\rLocal$, 
 four local distinct variables are created (say $h1,h2,l1,l2$), 
 the store becomes $l_a=[h1 | l1] \sqcup l_b=[h2 | l2]\sqcup c(h1,h2) $ 
 and the process $p(l1,l2)$ is executed on this new store. 
\end{itemize}
These  two \ccp\  executions match exactly the   behavior of the  
CLP goal \texttt{p(LA, LB)}.

We emphasize that one 
execution of a \ccp\ program will give rise to a single computation (due to the kind 
of non-determinism in \ccp) while the CLP 
abstract computation model characterizes the set of all possible 
successful derivations and corresponding answers. In other terms, 
for a given initial goal $G$, the CLP model defines the full set of
answer constraints for $G$, while the CCP translation will compute 
only one of them, as only one possible derivation will be followed.

\begin{theorem}[Adequacy]
Let $\cC$ be a constraint system, $c\in \cC$,  $\mathcal{H}$ be a CLP 
program and $G$ be a goal. Then, $\BarbCLP{ G  }{c} $ 
iff $\Barb{\os G\cs}{c}$. 
\end{theorem}

 \section{Slicing  CCP and CLP programs}\label{sectionslicing}

Dynamic slicing is a technique that helps the user to debug her program  by 
simplifying a partial execution trace, thus depurating it from parts which
are irrelevant to find the bug. 
It can also help to highlight parts of the programs which have 
been wrongly ignored by the execution of a wrong piece of code.
In \cite{FGOP2016} we defined a slicing technique for \ccp\ programs that consisted of three main steps:
\begin{enumerate}
 \item[{\bf S1}] \emph{Generating a (finite) trace} 
 of the program. For that,  a new semantics  is needed in order to generate  the (meta) information needed for the slicer. 
 \item[{\bf S2}] \emph{Marking the final store}, to select some of the constraints that, 
 according to the wrong behavior detected, should or should not be in the final store. 
 \item[{\bf S3}] \emph{Computing the trace slice}, to select the processes and constraints 
 that were relevant to produce the (marked) final store. 
\end{enumerate}

We shall briefly  recall the step  {\bf S1}  in  \cite{FGOP2016}  which remains the same here.  
Steps  {\bf S2} and   {\bf S3} need  further adjustments to deal with CLP programs. 
In particular, we shall allow the user to select processes (literals in the CLP terminology) 
in order to start  the slicing. Moreover, in Section  
\ref{sectionassertions}, we provide further tools to automatize the 
slicing process.

\paragraph{\bf Enriched Semantics (Step ${\bf S1}$).}\label{subsec:collect}
The slicing process  requires some extra information  from the execution of the processes. More precisely,    (1) in each operational step $\gamma \to \gamma'$, we need to highlight the process that was reduced; and (2) the constraints accumulated in the store must reflect, exactly, the contribution of each process to the store.  
In order to solve (1) and (2), we introduced 
in \cite{FGOP2016} the enriched semantics
 that extracts the needed meta information   for the slicer. 
Roughly, we identify  the  parallel composition 
 $Q = P_1 \parallel  \cdots \parallel P_n$ 
 with the   \emph{sequence}  $\Gamma_{Q} =P_1\idxP{i_1}, \cdots, P_n\idxP{i_n}$
 where $i_j   \in \mathbb{N}$ is a unique identifier for $P_j$.  
The use of indexes  allow us to distinguish, e.g.,   the three different  occurrences of $P$ in   
  ``$\Gamma_1,P\idxP{i},\Gamma_2,P\idxP{j}, (\whenp{c}{P})\idxP{k}$''. 
The enriched semantics  uses transitions with labels  of the form  $\rediIdxJ{i}{k}$ where $i$ is the identifier of the reduced process  and $k$ can be either 
$\bot$ (undefined) or a natural number indicating the branch chosen in a non-deterministic 
choice (Rule $\rSumCS$). This allows us to identify, unequivocally, the selected alternative in an execution. 
Finally, the  
store in the enriched semantics is not a constraint 
(as in Figure \ref{fig:sos}) but   a set  of (atomic) constraints 
where  $\{d_1,\cdots,d_n\}$
  represents the store 
$d_1 \sqcup \cdots \sqcup d_n$. 
For that, the rule of $\tellp{c}$  first   decomposes $c$ in its atomic components before  adding them to the store.

\paragraph{\bf Marking the Store (Step ${\bf S2}$).}\label{sec:step2}
In \cite{FGOP2016} we identified several alternatives for marking the final  store 
in order to indicate the information that is 
relevant to the slice that the programmer wants to recompute.
Let us suppose that the final configuration in a partial computation is $(X;\Gamma;S)$. 
The user has to select a subset $S_{sliced} $ of the final store $S$ that may explain the 
(wrong) behavior of the program.  $S_{sliced} $ can be 
chosen based on the following criteria: 

\begin{enumerate}
\item \emph{Causality:} the user
identifies, according to her knowledge, a subset $S' \subseteq S$ that needs to be explained 
(i.e., we need to identify the processes that produced $S'$). 
\item \emph{Variable Dependencies:} The user may identify a set of relevant variables 
$V\subseteq freeVars(S)$ and then, we mark 
$
S_{sliced} = \{ c \in S \mid vars(c) \cap V \neq \emptyset \}
$. 

 \item \emph{Unexpected behaviors}: there is a constraint $c$ entailed from the final 
 store that is not expected  from the  intended behavior of the program. 
 Then, one would be interested in the following marking 
$
S_{sliced} = \bigcup \{S' \subseteq S \mid  \bigsqcup S' \entails c 
\mbox{ and } S' \mbox{ is set minimal} \}
$, 
where ``$S'$ is set minimal'' means that for any $S'' \subset S'$, $S'' \not\entails c$. 
 \item \emph{Inconsistent output}: The final store should be consistent with respect  
 to a given specification (constraint) $c$, i.e., $S$ in conjunction with $c$ must not 
 be inconsistent. In this case, we have $
 S_{sliced} = \bigcup \{S' \subseteq S \mid \bigsqcup S' \sqcup c \entails \false 
\mbox{ and } S' \mbox{ is set minimal} \}
$. 
 \end{enumerate}

For the analysis of CLP programs, it is important also to mark literals (i.e., calls to procedures in \ccp). In particular, the programmer may find that a particular goal $p(x)$  is not correct if the parameter $x$ does not satisfy certain  conditions/constraints. Hence, we shall consider also markings on the set of processes, i.e., 
the marking can be also a subset  $\Gamma_{sliced} \subseteq \Gamma$.

\paragraph{\bf Trace Slice (Step ${\bf S3}$). }
Starting from the the pair  $\gamma_{sliced}= (S_{sliced}, \Gamma_{sliced})$
denoting the user's marking, we define 
a backward slicing step.
Roughly, this step  allows us to eliminate from the execution trace all the information  not related  to $\gamma_{sliced}$. 
For that, the fresh constant symbol $\sliced$ 
is used to denote an ``irrelevant'' constraint or process. 
Then, for instance, ``$c\sqcup \sliced$'' results from a constraint 
$c\sqcup d$ where $d$ is irrelevant.  Similarly in processes as, e.g., 
$\whenp{c}{(P \parallel \sliced)} + \sliced$. A replacement is either a pair of the shape $[T / i]$  or $[T / c]$. In the first (resp. second) case, 
 the process with identifier $i$  (resp. constraint $c$) is 
replaced with  $T$.  We shall use $\theta$ to denote a set of replacements
and we  call these sets  as 
``replacing substitutions''.  The  composition of  replacing substitutions $\theta_1$ and $\theta_2$ 
is given by the set union of   $\theta_1$ and 
$\theta_2$,
and is denoted as $\theta_1 \circ \theta_2$.

\begin{algorithm}[t]
{\scriptsize
\KwIn{- a trace $\gamma_0\rediIdxJ{i_1}{k_1} \cdots
\rediIdxJ{i_{n}}{k_{n}}\gamma_n$ 
where $\gamma_i= (X_i;\Gamma_i; S_i)$

  \qquad\quad\ - a marking ($S_{sliced} , \Gamma_{sliced}$) s.t. $S_{sliced} \subseteq S_n$ and $\Gamma_{sliced} \subseteq \Gamma_n$}
    
\KwOut{ a sliced trace $\gamma_0'\redi    \cdots \redi \gamma_n' $}

\Begin{
 {\bf let} $\theta = \{ [\sliced/i] \mid P\idxP{i} \in \Gamma_n \setminus \Gamma_s \}$ {\bf in}

 $\gamma_n' \leftarrow ( X_n \cap vars(S_{sliced}, \Gamma_{sliced}) ; \Gamma_{n}\theta; S_{sliced})$\;
 \For{l= $n-1$ to 0}{
 ${\bf let} \langle \theta', c\rangle  =  
 sliceProcess(\gamma_l, \gamma_{l+1}, i_{l+1}, k_{l+1},\theta, S_l)  \ 
 $ {\bf in}

$S_{sliced}  \leftarrow  S_{sliced} \cup  S_{minimal}(S_{l},c) $

$\theta  \leftarrow \theta' \circ \theta$
 
  $\gamma_l' \leftarrow (X_l\cap vars(S_{sliced}, \Gamma_{sliced}) ~; ~ \Gamma_l \theta ~;~ S_l \cap S_{sliced})$

  } 
}
}
 \caption{Trace Slicer.  $S_{minimal}(S,c)=\emptyset$ if $c=\true$; otherwise,  $S_{minimal}(S,c) = \bigcup \{S' \subseteq S \mid  \bigsqcup S' \entails c 
\mbox{ and } S' \mbox{ is set minimal} \}$. 
 \label{alg:slicer}}
\end{algorithm}

 Algorithm  \ref{alg:slicer} extends the one 
 in \cite{FGOP2016} to deal with the marking on processes ($\Gamma_{sliced}$).   The last configuration
 ($\gamma_n'$ in line 3) means that we only observe the local variables of interest, 
 i.e., those in $vars(S_{sliced}, \Gamma_{sliced})$ as well as the relevant processes ($\Gamma_{sliced}$)
 and constraints $(S_{sliced})$. 
The algorithm backwardly computes the slicing by accumulating 
replacing pairs in $\theta$ (line 7). The new replacing substitutions are computed by the function 
$sliceProcess$ 
that returns both, a replacement substitution and a constraint needed in the case  of ask agents as  explained below.

\begin{algorithm}[]
{\scriptsize
\SetKwProg{Fn}{Function}{ }{end}
\Fn{sliceProcess($\gamma, \psi, i,k, \theta, S$) }{
  {\bf let} $\gamma=(X_\gamma ; \Gamma,P\idxP{i},\Gamma'; S_\gamma)$ and $\psi=(X_\psi ; \Gamma,\Gamma_Q, \Gamma' ; S_\psi)$ {\bf in}
    \SetKw{KMatch}{match}
  \SetKw{KWith}{with}
  \SetKwBlock{KBMatch}{\KMatch{$P$ \KWith }}{end}

  \KBMatch{
     \uCase{$\tellp{c}$}{
     {\bf let} $c' = sliceConstraints(X_\gamma,X_\psi, S_\gamma, S_\psi, S)$ {\bf in}
     
       \leIf{$c' = \sliced$ or $c' = \exists \overline{x}. \sliced$}{\KwRet{ $\lr{[\sliced / i] , \true}$}}{\KwRet{ $\lr{[\tellp{c'}/i],\true}$}}
       
     } 
     \uCase{$\sum\whenp{c_l}{Q_l}$}{
     \leIf{$ \Gamma_Q \theta=\sliced$}{\KwRet{ $\lr{[\sliced / i],\true}$}}{\KwRet{ 
     $\lr{[ \whenp{c_k}{(\Gamma_Q\theta)} + \sliced ~/~ i], c_k}$}}
     } 
     \uCase{$\localp{x}{Q}$}{
     {\bf let} $\{x'\} = X_\psi \setminus X_\gamma$ {\bf in}
     
       \leIf{$\Gamma_Q[x'/x] \theta=\sliced$}{\KwRet{ $\lr{[\sliced / i],\true}$}}{\KwRet{ $\lr{[\localp{x'}{\Gamma_Q[x'/x]\theta}/ i],\true}$}}
     } 
     \uCase(){$p(\vy)$}{
     
     	\leIf{$\Gamma_Q \theta=\sliced$}{\KwRet{ $\lr{[\sliced / i],\true}$}}{\KwRet{ $\lr{\emptyset, \true}$}}

     } 
  }
  
} 

\SetKwProg{Fn}{Function}{ }{end}
\Fn{sliceConstraints($X_\gamma, X_\psi, S_\gamma, S_\psi, S$) }{
{\bf let} $S_c = S_\psi \setminus S_\gamma \mbox{ and } \theta= \emptyset$ {\bf in}

 \lForEach{$c_a \in S_c \setminus S$}{
   $\theta \leftarrow \theta \circ [\sliced / c_a]$
 }
 
 \KwRet{ $\exists_ {X_{\psi} \setminus X_{\gamma}}. \bigsqcup  S_c \theta$}
}
}
 \caption{Slicing processes and constraints \label{alg:proc}}
\end{algorithm}

\noindent
{\bf Marking algorithms. }
 Let us explain how the function $sliceProcess$  works. 
Consider for instance 
the process $Q=(\whenp{c'}{P})+(\whenp{c}{\tellp{d\sqcup e}}) $
and assume that we are backwardly slicing the trace   $\cdots \gamma \rediIdxJ{i}{2}  \cdots \psi \rediIdxJ{j}{} \rho \cdots $
where $Q$ (identified with $i$) is reduced in $\gamma$ by choosing  the second branch  and, in $\psi$, 
the tell agent $\tellp{d\sqcup e}$ (identified by $j$) is executed.  
Assume that the configuration $\rho$ has already been sliced and
$d$ was considered irrelevant and removed (see $S_l \cap S_{sliced}$ in line 8 of Algorithm \ref{alg:slicer}). 
The procedure $sliceProcess$ is applied to $\psi$ and it determines that 
only $e$ is relevant in $\tellp{d\sqcup e}$.  Hence, 
the replacement $[\tellp{\sliced \sqcup e}/j]$ is returned (see line 7 in 
Algorithm \ref{alg:slicer}). The procedure is then applied to $\gamma$. We already know 
that the ask agent $Q$ is (partially) relevant since $\tellp{d\sqcup e}\theta \neq \sliced$ 
(i.e., the selected branch does contribute to the final result).  Thus, the replacement 
$[\sliced + \whenp{c}{\tellp{\sliced\sqcup e}} / i]$ is accumulated in order to show 
that the first branch is irrelevant. Moreover, 
since the entailment of $c$ was necessary for the reduction,  
the procedure returns also the constraint $c$ (line 5 of Algorithm \ref{alg:slicer}) and 
the constraints needed to entail $c$ are added to the set of relevant constraints 
(line 6 of   Algorithm \ref{alg:slicer}).

\begin{example}\label{ex:length}
Consider the following (wrong) CLP program: 
\begin{Verbatim}[fontsize=\scriptsize]
length([],0).
length([A | L],M) :- M = N, length(L, N).
\end{Verbatim}
The translation to \ccp\ is similar to the example we gave in Section \ref{sec:clp}. 
An excerpt of a possible trace for the execution of the goal \verb|length([10,20], Ans).| is
\begin{Verbatim}[fontsize=\tiny]
[0 ; length([10,20],Ans) ; t] --> 
[0 ; ask() ... + ask() ... ; t] -> 
[0 ; local ... ; t] -> 
[H1 L1 N1 M1 ; [10,20]= [H1|L1] || Ans=N1 || N1=M1 || length(L1, M1) ; t] ->
...
[... H2 L2 N2 M2 ; [20]=[H2 | L2] || M1=N2 || N2=M2 || length(L2, M2) ; [10,20]= [H1|L1], Ans=N1, N1=M1] ->
[... H2 L2 N2 M2 ; M1=N2 || N2=M2 || length(L2, M2) ; [10,20]= [H1|L1], Ans=N1, N1=M1, [20]=[H2 | L2]] -> 
...
[... H2 L2 N2 M2 ; M2=0 ; [10,20]= [H1|L1], Ans=N1, N1=M1, [20]=[H2 | L2], M1=N2, N2=M2, L2=[]] ->  
[... H2 L2 N2 M2 ; [10,20]= [H1|L1], Ans=N1, N1=M1, [20]=[H2 | L2], M1=N2, N2=M2, L2=[], M2=0 ] 
\end{Verbatim}
In this trace, we can see how the calls to 
the process definition \texttt{length} are unfolded 
and, in each state, new constraints are added. 
Those constraint relate, e.g., the variable \texttt{Ans} 
and the local variables created in each invocation  
(e.g., \texttt{M1} and \texttt{M2}). 

In the last configuration, it is possible to  mark only the equalities dealing with numerical expressions (i.e., 
\verb|Ans=N1,N1=M1,M1=N2,N2=M2,M2=0|) and the resulting trace will abstract away from all the constraints and processes dealing with equalities on lists:
\begin{Verbatim}[fontsize=\tiny]
[0 ; length([10,20],Ans) ; t] --> 
[0 ; * + ask() ... ; t] -> 
[0 ; local ... ; t] -> 
[N1 M1 ; * || Ans=N1 || N1=M1 || length(L1, M1) ; t] ->
[N1 M1 ; Ans=N1 || N1=M1 || length(L1, M1) ; ] -> 
[N1 M1 ; N1=M1 || length(L1, M1) ;   Ans=N1] ->
[N1 M1 ; length(L1, M1) ;  Ans=N1, N1=M1] -> 
...
\end{Verbatim}
The fourth line should be useful to discover that \verb|Ans| cannot be equal to  \verb|M1| (the parameter used in the second invocation to \verb|length|). 

\end{example}

 \section{An assertion language for logic programs}\label{sectionassertions}

The declarative flavor of programming with constraints in \ccp\ and CLP 
allows the user to reason about (partial) invariants that must hold during the 
execution of her programs. In this section we give a simple yet powerful 
language of assertion to state such invariants. Then, we give a step further 
in automatizing the process of debugging. 

\begin{definition}[Assertion Language]\label{sec:syntax}
Assertions are built from the following syntax: 

\noindent\resizebox{.95\textwidth}{!}{
$
F ::= \posC{c}  ~\mid~ \negC{c} ~\mid~ \consC{c} ~\mid~ \iconsC{c} ~\mid~  F \oplus F   ~\mid~ \predAssertionA{p(\vx)}{F} ~\mid~ \predAssertionE{p(\vx)}{F}  
$}

\noindent where $c$ is a constraint ($c\in \mathcal{C}$), $ p(\cdot) $ is a process name and $\oplus \in \{\wedge, \vee, \to \}$. 
\end{definition}

The first four constructs deal with partial assertions about the current store. These constructs check, respectively, whether the constraint $c$:  (1) is entailed, (2) is not entailed, (3) is consistent wrt the current store or (4)  leads to an inconsistency when added to the current store. 
Assertions of the form $F \oplus F$ have the usual meaning. 
The assertions $\predAssertionA{p(\vx)}{F}$
 states that all instances 
 of the form    $p(\vt)$ in the current configuration  must satisfy the assertion $F$. 
The assertions $\predAssertionE{p(\vx)}{F}$ is similar to the previous one but  it checks for the existence 
of an instance $p(\vt)$ that satisfies  the assertion $F$.

Let $\pi(i) =   (X_i ; \Gamma_i ; S_i)$. We shall use $\fStore{\pi(i)}$ to denote the constraint $\exists X_i . \bigsqcup S_i$ and $\fProc{\pi(i)}$ to denote the sequence of processes $\Gamma_i$. The semantics for assertions is formalized next.

\begin{definition}[Semantics]\label{sec:semantics}
Let $\pi$ be a sequence of configurations and $F$ be an assertion.  
We inductively define $\pi,i \entailsF F$ (read as  $\pi$ satisfies the formula $F$ at position $i$) as: 
\begin{itemize}
 \item   $\pi, i \entailsF  \posC{c}$ if $store(\pi(i)) \entails c$. 
 \item   $\pi, i \entailsF  \negC{c} $ if $store(\pi(i)) \not\entails c$. 
 \item   $\pi, i \entailsF  \consC{c} $ if $store(\pi(i)) \sqcup c \not\entails \false$. 
  \item   $\pi, i \entailsF  \iconsC{c} $ if $store(\pi(i)) \sqcup c \entails \false$. 
 \item   $\pi, i \entailsF F \wedge G$ if $\pi,i \entailsF F$ and $\pi,i \entailsF G$. 
 \item   $\pi, i \entailsF F \vee G$ if $\pi,i \entailsF F$ or $\pi,i \entailsF G$. 
  \item   $\pi, i \entailsF F \to G$ if $\pi,i \entailsF F$ implies $\pi,i \entailsF G$. 
 \item  $\pi, i \entailsF  \predAssertionA{p(\vx)}{F} $ if for all $p(\vt)\in \fProc{\pi(i)}$, $\pi, i \entailsF F[\vt/\/\vx]$.
 \item  $\pi, i \entailsF  \predAssertionE{p(\vx)}{F} $ if there exists $p(\vt)\in \fProc{\pi(i)}$, $\pi, i \entailsF F[\vt/\/\vx]$.

\end{itemize}
If it is not the case that $\pi, i \entailsF  F $, then we say that $F$ does not hold at $\pi(i)$ and we write $\pi(i) \not\entailsF F$. 
\end{definition}

The above definition  is quite standard and reflects  the intuitions given above.
Moreover, let us define $\sim F$ as $\sim \posC{c} = \negC{c}$ (and vice-versa), $\sim \consC{c} = \iconsC{c}$ (and vice-versa), $\sim (F \oplus F)$ as usual
and $\sim \predAssertionA{p(\vx)}{F(\vx)} = \predAssertionE{p(\vx)}{\sim F(\vx)}$ (and vice-versa). Note that, 
$\pi(i) \entailsF F$ iff $\pi(i) \not\entailsF \sim F$.

\begin{example}
Assume  that the  store in $\pi(1)$ is  $S = x \in 0..10$. Then,

\noindent - ${\pi,1 \entailsF \consC{x=5}}$, i.e., the current store is consistent wrt the specification $x=5$. 

\noindent - $\pi,1 \not\entailsF \iconsC{x=5}$, i.e.,  the   store is not inconsistent wrt the specification $x=5$. 

\noindent - $\pi,1 \not\entailsF \posC{x=5}$, i.e.,  the store is not ``strong enough''  in order to satisfy the specification $x=5$. 

\noindent - $\pi,1 \entailsF \negC{x=5}$, i.e.,  store is ``consistent enough'' to guarantee that  it is not the case that  $x=5$.

\end{example}
Note that $\pi,i \entailsF \posC{c}$  implies   $ \pi,i \entailsF \consC{c}$. However, the other direction is in general not true (as shown above).
We note that \ccp\ and CLP are monotonic in the sense that
when the store $c$ evolves into $d$, it must be the case that $d \entails c$ (i.e., information is  monotonically accumulated).  Hence,   $\pi,i \entails \posC{c}$ implies $\pi,i + j \entails \posC{c}$. Finally, if the store becomes inconsistent,  $\consC{c}$ does not hold for any $c$. 
Temporal \cite{NPV02} and linear \cite{fages01ic} variants of   \ccp\ 
remove such restriction on monotonicity.

We note that checking assertions amounts, roughly, to testing the entailment 
relation in the underlying constraint system. Checking entailments is the basic 
operation \ccp\ agents perform. Hence, from the implementation point of view, 
 verification of assertions 
does not introduce a significant extra  computational cost. 

\begin{example}[Conditional assertions]\label{ex:patterns}
Let us introduce some patterns of assertions useful for verification. 

\noindent- \emph{Conditional constraints} : 
The assertion $\posC{c} \to F$ checks for $F$ only if $c$ 
can be deduced from the store. 
For instance, the assertion $\posC{c} \to \negC{d}$ says that $d$ 
must not be deduced when the store implies $c$.

\noindent- \emph{Conditional predicates} :
Let $G = \predAssertionE{p(\vx)}{\consC{\true}}$. 
The assertion $G\to F$
states that $F$ must be verified  whenever there is a call/goal of the form $p(\vt)$ in the context. Moreover, $ (\sim G) \to F $
verifies $F$ when there are no calls of the form $p(\vt)$ in the context. 
\end{example}

\subsection{Dynamic slicing with assertions}
Assertions allow the user to specify conditions that her program must 
satisfy during execution. If this is not the case, the program should stop 
and start the debugging process. In fact, the assertions may help to give 
a suitable marking pair $(S_{sliced}, \Gamma_{sliced})$ for the step 
${\bf S2}$  of our algorithm as we show in the next definition.

\begin{definition}\label{def:symp}
Let $F$ be an assertion, $\pi$
be a partial computation, $n>0$  and assume that 
$\pi, n \not\entailsF F$, i.e., $\pi(n)$ fails to establish the assertion $F$. 
Let     $\pi(n ) = (X  ; \Gamma  ; S )$. As testing hypotheses, we define
$\fSymp{\pi,F, n} = (S_{sl}, \Gamma_{sl})$ where

\begin{enumerate}
\item If $F = \posC{c}$ then $S_{sl}=\{d \in S \mid vars(d)\cap vars(c) \neq \emptyset \}$, $\Gamma_{sl}=\emptyset$. 
\item If $F = \negC{c}$ then $S_{sl}=\bigcup \{S' \subseteq S \mid  \bigsqcup S' \entails c 
\mbox{ and } S' \mbox{ is set minimal} \}$, $\Gamma_{sl}=\emptyset$
\item If $F = \consC{c}$ then $S_{sl}=\bigcup \{S' \subseteq S \mid \bigsqcup S' \sqcup c \entails \false 
\mbox{ and } S' \mbox{ is set minimal} \}$, $\Gamma_{sl}=\emptyset$. 
\item If $F = \iconsC{c}$ $S_{sl}=\{d \in S \mid vars(d)\cap vars(c) \neq \emptyset \}$ and $\Gamma_{sl}=\emptyset$. 

\item If $F = F_1 \wedge F_2$ then $\fSymp{\pi, F_1, n } \cup \fSymp{\pi, F_2, n }$. 
\item If $F = F_1 \vee F_2$ then $\fSymp{\pi, F_1, n } \cap \fSymp{\pi, F_2, n }$. 
\item If $F = F_1 \to F_2$ then $\fSymp{\pi, \sim F_1, n } \cup \fSymp{\pi, F_2, n }$. 
\item If $F = \predAssertionA{p(\vx)}{F_1}$ then $S_{sl}=\emptyset$ and $\Gamma_{sl}= \{p(\vt) \in \Gamma \mid \pi,n \not\entailsF F_1[\vt/\vx]\}$. 
\item If $F = \predAssertionE{p(\vx)}{F_1}$ then $S_{sl}=\{d\in S \mid vars(d) \cap vars(F_1) \neq \emptyset\}$, $\Gamma_{sl}=\{p(\vt) \in \Gamma\}$

\end{enumerate}

\end{definition}

Let us give some intuitions about the above definition. 
Consider a (partial) computation $\pi$ of length $n$ where $\pi(n) \not\entailsF F$. 
In the  case (1) above, $c$ must be entailed but the current store is not strong enough to do it. A good guess is to start examining the processes that added constraints using the same variables as in $c$. 
It may be the case that such processes should have added more information to entail $c$ as expected in the specification $F$. 
Similarly for the case (4): $c$ in conjunction with the current store should be inconsistent but it is not. Then, more information on the common variables should have been added. 
In the case (2), $c$ should not be entailed but the store indeed entails $c$. In this case, we mark the set of constraints that entails $c$. The case (3) is similar. In cases (5) to (7) we use $\cup$ and $\cap$ respectively for point-wise union and intersection in the pair $(S_{sl} , \Gamma_{sl})$. 
These cases are self-explanatory 
(e.g., if $F_1\wedge F_2$ fails, we collect the failure information 
of  either   $F_1$ or  $F_2$). 
In (8), we mark all the calls that do not satisfy the expected assertion $F(\vx)$. In (9), if $F$ fails, it  means that
either (a) there are no calls of the shape $p(\vt)$ in the context or (b) none of the calls $p(\vt)$ satisfy $F_1$. For (a),  
similarly to the case (1), a good guess is to examine the processes that added constraints with common variables to $F_1$ and  see which one should have added more information to entail $F_1$. As for (b), we also select all the calls of the form $p(\vt)$ from the context. 
The reader may compare these definitions with the information selected
in Step {\bf S2} in Section \ref{sectionslicing}, regarding possibly wrong 
behavior. 

\paragraph{Classification of Assertions. } 
As we explained in Section \ref{sec:clp}, computations in CLP 
can succeed or fail and the answers to a goal is the set of constraints 
obtained from successful computations. Hence, according to the kind 
of assertion, 
it is important to
determine when the assertions in Definition \ref{sec:syntax}
  must  stop  or not the computation to start the debugging process. For that, we  introduce the following classification: 

\noindent  {\bf - post-conditions, $\postF{F}$ assertions }: assertions that are meant to be verified only when an answer is found. 
This kind of assertions are used to test the ``quality'' of the answers wrt the specification.  In this case, 
 the slicing process begins only when an answer is computed and it does not satisfy one of the assertions.  
 Note that assertions of the form 
 $\predAssertionA{p(\vx)}{F(\vx)}$ and  $\predAssertionE{p(\vx)}{F(\vx)}$ are irrelevant as post-conditions since the 
set of goals in an answer must be empty.

\noindent  {\bf - path invariants, $\invF{F}$ assertions}: assertions that are meant to hold along the whole computation. Then, not satisfying an invariant must be understood as a 
 symptom of an error and the computation must stop. 
We note that due to monotonicity, 
 only assertions of the form $\negC{c}$ and $\consC{c}$ can be used to stop the computation (note that if the current configuration fails to satisfy  $\negC{c}$, then any successor state will also fail to satisfy that assertion). 
 Constraints of the form $\posC{c}, \iconsC{c}$  can be only checked 
 when the answer is found since, not satisfying those conditions in 
 the partial computation, does not imply that the final state will not satisfy them.

\subsection{Experiments}\label{sec:ex}
We conclude this section with a series of examples   showing the use of assertions. Examples \ref{test1} and \ref{test2} 
deal with CLP programs while Examples \ref{test3} and \ref{test4} with \ccp\ programs. 

\begin{example}\label{test1}
The debugger can automatically start and produce the same marking in  Example \ref{ex:length} with the following (invariant) assertion: 
\begin{Verbatim}[fontsize=\scriptsize]
length([A | L],M) :- M = N, length(L, N), inv(pos(M>0)).
\end{Verbatim}
\end{example}

\begin{example}\label{test2}
Consider the following CLP 
program (written in GNU-Prolog with integer finite domains) for solving the well known 
problem of posing $N$ queens on a $N\times N$ chessboard 
in such a way that they do not attack each other. 
\begin{Verbatim}[fontsize=\scriptsize]
queens(N, Queens) :- length(Queens, N), fd_domain(Queens,1,N),
                     constrain(Queens), fd_labeling(Queens,[]).
constrain(Queens) :-fd_all_different(Queens), diagonal(Queens).
diagonal([]).
diagonal([Q|Queens]):-secure(Q, 1, Queens), diagonal(Queens).
secure(_,_,[]).
secure(X,D,[Q|Queens]) :- doesnotattack(X,Q,D),D1 is D+1, secure(X,D1,Queens).
doesnotattack(X,Y,D) :- X + D #\= Y,Y + X #\= D.
\end{Verbatim}

\noindent
The program  contains one mistake, which causes the introduction of 
a few additional and not correct solutions, e.g., \verb|[1,5,4,3,2]| 
for the goal  \verb|queens(5,X).| 
 The user now has two possible strategies:  either she lets the 
interpreter compute the solutions, one by one and then,  when she sees a wrong  solution  she uses the slicer for marking manually the 
final store to  get  the sliced computation; or she can define an assertion to be verified. 
In this particular case, any solution must satisfy that the difference
between two consecutive positions in the list must be greater than $1$.
Hence, the user can introduce the following   post-condition assertion:  

\begin{Verbatim}[fontsize=\scriptsize]
secure(X,D,[Q|Queens]) :- doesnotattack(X,Q,D),D1 is D+1, secure(X,D1,Queens),  
                          post(cons(Q #\= X+1)).   
\end{Verbatim}

Now the slicer stops as soon as the constraint \verb|X #\= Q+1|
becomes inconsistent with the store in a successful computation 
(e.g., the assertion fails on the --partial-- assignment ``5,4'')  
and an automatic slicing   is performed.

\end{example}

\begin{example}\label{test3}
In \cite{FGOP2016} we presented a compelling example of slicing for a timed \ccp\ program modeling the synchronization of events in musical rhythmic patterns. As shown in Example 2 at \url{http://subsell.logic.at/slicer/}, the slicer for \ccp\ was able to sufficiently abstract away from irrelevant processes and constraints to highlight the problem in a faulty program. However, the process of stopping the computation to start the debugging was left to the user. The property that failed in the program can be naturally 
expressed as an assertion. Namely, in the whole computation, if the constraint $\beatc$ is present (representing a sound in the musical rhythm), the constraint $\stopc$ cannot be present (representing the end of the rhythm). This can be written as the  conditional assertion   $\posC{\beatc} \to \negC{\stopc}$. Following Definition \ref{def:symp}, the constraints marked in the wrong computation are the same we considered in \cite{FGOP2016},   thus automatizing completely the process of identifying the wrong computation. 
\end{example}

\begin{example}\label{test4}
Example 3 in the  URL above illustrates the use of timed \ccp\ for the 
specification of biochemical systems (we invite the reader to compare in the 
website  the sliced and non-sliced traces). Roughly, in that model, constraints 
of the form $\Mdmtwo$ (resp. $\MdmtwoA$) state that the 
protein $\Mdmtwo$ is present (resp. absent). The model includes activation 
(and inhibition)  of biological rules modeled  as processes (omitting some 
details) of the form 
$\whenp{\MdmtwoA}{\nextp{\tellp{\Mdmtwo}}}$ modeling  that ``if
 $\Mdmtwo$ is absent now, then  it must be present in the next time-unit''. 
The interaction of many of these rules makes the model trickier  since  rules 
may ``compete'' for resources and then, we can wrongly observe at the same 
time-unit that $\Mdmtwo$ is both present and absent.
An assertion of the form 
$(\posC{\MdmtwoA} \to \negC{\Mdmtwo})\wedge(\posC{\Mdmtwo} \to \negC{\MdmtwoA})$
will automatically stop the computation and  produce the same marking we 
used to 
depurate the program in  the website. 
\end{example}

 \section{Related work and conclusions}\label{sectionconclusions}

{\bf Related work} Assertions for automatizing a slicing process have been 
previously
introduced in \cite{Alpuente2016DebuggingMP} for the functional logic
language Maude. The language they consider as well as the 
type of assertions are completely different from ours. 
They do not have constraints, and deal with functional and equational
computations. Another previous work \cite{SGM2002}
introduced static and dynamic 
slicing for CLP programs. However, \cite{SGM2002} essentially aims at
identifying the parts of a goal which do not share variables, to divide
the program in slices which do not interact. 
Our approach considers more situations, not only
variable dependencies, but also other kinds of wrong behaviors.
Moreover we have assertions, and hence an automatic 
slicing mechanism not considered in \cite{SGM2002}. 
The well known debugging box model  of Prolog \cite{ClocksinMellish1981} 
introduces 
a tool for observing the evolution of atoms during their reduction in the search 
tree. We believe that our methodology might be integrated with the box model 
and may extend some of its features. For instance, the box model makes 
basic simplifications by asking the user to specify which predicates she 
wants to observe. In our case, one entire computational path is simplified 
automatically by considering the
marked information and identifying the constraints and the 
atoms which are relevant for such information.

\noindent
{\bf Conclusions and future work} In this paper we have first extended a previous framework
for dynamic slicing of (timed) \ccp\ programs to the case of CLP
programs. We considered a slightly different marking
mechanism, extended to atoms besides constraints.
Don't know non-determinism in CLP requires a different
identification of the computations of interest  wrt CCP. 
We considered different modalities specified by the
user for selecting successful computations rather than all
possible partial computations.  
As another contribution of this paper,
in order to automatize the slicing process, we 
have introduced an assertion
language. This language is rather flexible and allows one to specify 
different types of assertions that can be applied 
to successful computations or to all possible partial computations.
When assertions are not satisfied by
a state of a selected computation then an automatic slicing
of such computation can start.

We implemented a  prototype of the slicer in Maude
and showed its use in debugging several programs. 
We are currently extending the tool to deal with CLP
don't know non-determinism.  Being CLP a generalization  of logic programming, 
our extended implementation could be also eventually used to analyze Prolog 
programs.  Integrating the kind of assertions proposed here with already 
implemented debugging mechanisms in Prolog is an interesting future direction. 
We also plan to add more advanced graphical tools to
our prototype, as well as to study the integration of our
framework with other debugging techniques  such as 
the box model and declarative or approximated debuggers
\cite{FOPV07,ABCF10}. We also
want to investigate the relation of our technique with dynamic testing
(e.g. concolic techniques) and
extend the assertion language with temporal operators, e.g. the past 
operator ($\past$) for expressing the
relation between two consecutive states.
Another future topic of investigation is a static
version of our framework in order to try to compare and possibly
integrate it with analyses and semi automatic corrections
based on different formal techniques, and other programming paradigms
\cite{AFMV99,BBB09,BBBC14,ABFR06,ABBF10}.

\medskip

\noindent
{\small {\bf Acknowledgments}
We thank the anonymous reviewers for their detailed and very useful 
criticisms and recommendations that helped us to improve our paper. The  work of Olarte  was supported by CNPq and by CAPES,
Colciencias, and INRIA via the STIC AmSud project EPIC (Proc. No 
88881.117603/2016-01), and the project CLASSIC.}

\bibliographystyle{plain}

\end{document}